\documentclass[preprint]{aastex}

\begin{document}

\title{Polarimetry in the Visible and Infrared: Application to CMB Polarimetry}

\author{Terry J. Jones}
\author{Department of Astronomy, University of Minnesota, 116 Church St SE, Minneapolis, MN 55455}

\begin{abstract}

Interstellar polarization from aligned dust grains can be measured both in transmission at visible and near-infrared wavelengths and in emission at far-infrared and sub-mm wavelengths. These observations can help predict the behavior of foreground contamination of CMB polarimetry by dust in the Milky Way. Fractional polarization in emission from aligned dust grains will be at the higher range of currently observed values of 4-10\%. Away from the galactic plane, fluctuations in Q and U will be dominated by fluctuations in intensity, and less influenced by fluctuations in fractional polarization and position angle.

\end{abstract}

\section{Some Basics}

Polarization due to magnetically aligned dust grains in the Milky Way galaxy is subject to both coherent and random processes (see Jones, 1996 for a review of grain alignment). Most observations of interstellar polarization have been at visual wavelengths, with a modest contribution at near-infrared (NIR) wavelengths. At these wavelengths the polarization is due to selective extinction by aspherical dust grains. In the far-infrared (FIR), these same grains are in emission. This emission is polarized and consequently a foreground contaminant for CMB polarimetry. A few simple facts about interstellar polarization should be noted.

\begin{itemize}
   \item Fractional polarization P is very probably NOT a function of magnetic field strength \citep{jkd}.
   
   \item Polarization in emission (P$_e$) behaves differently than polarization in transmission (P$_a$).
   
   \item P$_a$ starts low, then increases with increasing optical depth and contains information on the random component of the aligning magnetic field.
   
   \item P$_e$ starts high, then decreases with increasing optical depth. CMB polarimetry will sample the low optical depth regime.
   
\end{itemize}

CMB polarization measurements will be concerned with contamination from fluctuations in Stokes vectors Q and U caused by variations in interstellar polarization in emission. Recall that Q and U are intensities, not fractions. Specifically:

\begin{eqnarray*}
I_{P} &=&P_{frac}\cdot I_{tot} \\
I_{P} &=&\sqrt{Q^{2}+U^{2}} \\
Q &=&I_{P}\cos (\theta ) \\
U &=&I_{P}\sin (\theta ) \\
\theta  &=&\frac{1}{2}\tan ^{-1}\left( \frac{U}{Q}\right) 
\end{eqnarray*}

\noindent Thus, Variations in Q and U can be due to fluctuations in fractional polarization, fluctuations in position angle, and fluctuations in total intensity. For example, a region of sky with the same fractional polarization in emission (uniform grain alignment across the sky) will still show fluctuations in Q and U if the intensity is varying across the region. At high latitudes, this effect will turn out to be the major source of contamination.

Polarimetry at FIR and sub-mm wavelengths would seem the best source of observations for characterizing the CMB polarized foreground. These observations measure polarized emission from grains, and in the sub-mm extend to wavelengths close to those to be observed in upcoming and proposed CMB polarimetry experiments. However, FIR and sub-mm observations require significant optical depth to achieve detectable emission, and consequently have largely been confined to dense regions in the galactic plane. At high latitudes, where CMB polarimetry will be most effective, the emission is very optically thin and nearly impossible to observe in polarization. The reader should see \cite{hil00} for a thorough discussion of FIR polarimetry.

\section{Observations of Interstellar Polarization}

\begin{figure}[!ht]
\begin{center}
\includegraphics[scale=0.8]{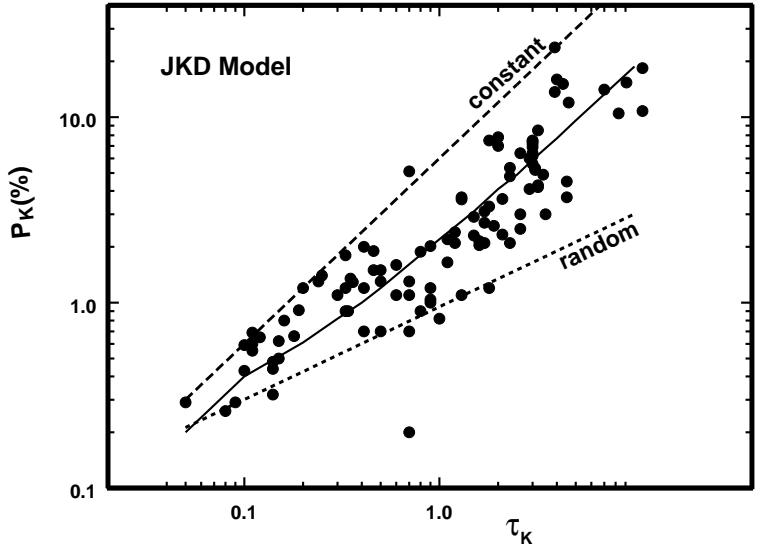}
\caption{\footnotesize Fractional polarization vs. optical depth at K (2.2~\micron) and the JKD model. The solid model line corresponds to a 50/50 mix of random and constant components to the interstellar magnetic field with a decorrelation length of $\tau_{K} = 0.1$}
\end{center}
\end{figure}

Optical polarimetry can, however, measure polarization in transmission (P$_a$) at high latitudes because dust extinction is so strongly wavelength dependent. We need to relate optical, NIR and FIR polarimetric observations in order to predict the characteristics of foreground polarized emission at high latitudes. This work is in its infancy, but some useful conclusions can be made.

\cite{jkd} present a simple model of grain alignment by the interstellar magnetic field that they use to explain the general trend of polarization (P$_a$) with optical depth ($\tau$) observed at optical and NIR wavelengths. They assume the magnetic field is strong enough everywhere to fully align the grains and that any departure from a simple linear relation between P$_a$ and $\tau$ is due to fluctuations in the magnetic field {\it geometry}. Their results are shown in Figure 1. The upper model line is the case for a perfectly uniform magnetic field geometry, the lower model line is for a totally random magnetic field geometry. The data clearly lie between these two extremes. The solid middle line is the model result with a 50/50 mix of random and uniform components and a decorrelation length for the random component of $\tau_{K} = 0.1$. 

Similar results have been obtained by \cite{fos02} using a much larger optical data set. These model results show that fluctuations in $P_a$ will be strongly influenced by fluctuations in the magnetic field geometry only for optical depths greater than $\tau \sim 0.03-0.1$ at 2.2~\micron\ (about $A_{V} = 0.3-1.0$). At high galactic latitudes, where the optical depth is much less than this, we would expect the magnetic geometry to be relatively smooth, and therefore not a significant factor influencing fluctuations in Q and U in emission at FIR wavelengths. 

This conclusion can be seen more clearly by using the parameters from the JDK model to predict the polarization in emission. The results of this exercise are shown in Figure 2, where the predicted fractional polarization at FIR wavelengths is plotted vs. the number of decorrelation cells along the line of sight. The model computes the predicted {\it average} polarization for an ensemble of lines of sight. Note that $P_{e}$ is highest at the low optical depths, since all of the grains are emitting with the same alignment geometry.  This is the regime in which CMB observations will be most useful. 

Based on the results from NIR polarimetry (Figure 1), one decorrelation cell corresponds roughly to $A_{V} = 0.3- 1$, or in physical distance units about 200-500 pc in the diffuse ISM. This corresponds to a greater optical depth than seen at high latitudes and one such cell would cover several degrees across the sky. Thus, the JKD model predicts that the CMB foreground contamination will most likely have a high intrinsic fractional polarization but a relatively uniform projected geometry over many degrees at high latitudes.

\begin{figure}[!ht]
\begin{center}
\includegraphics[scale=0.6]{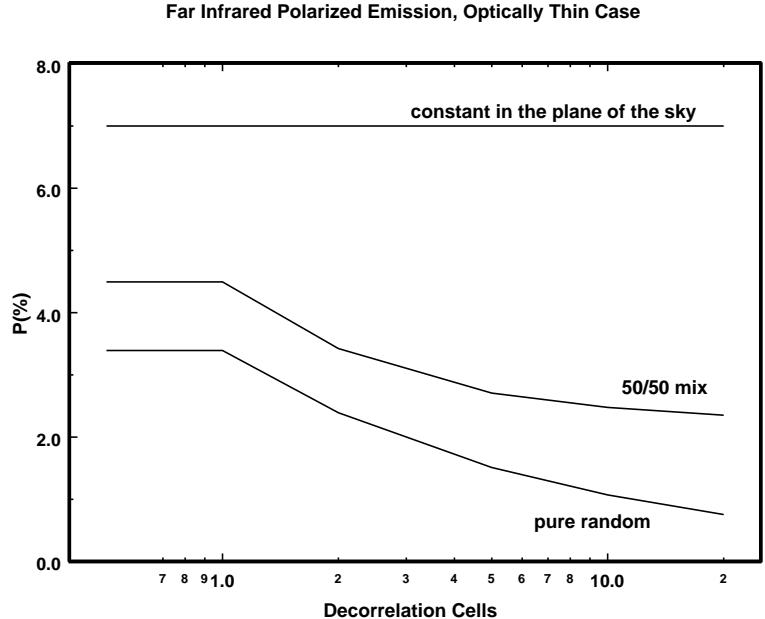}
\caption{\footnotesize Predicted {\it average} fractional polarization in the FIR vs. number of decorrelation cells using the JKD model. The middle line corresponds to a 50/50 mix of random and constant components. The polarization for this model is less than the maximum (7\%) due to the fact that the random component of the magnetic field can point in any direction, including along the line of sight.}
\end{center}
\end{figure}

In Figure 2 we have assumed the maximum polarization of completely aligned grains is 7\%, a tyical upper bound for FIR measurements in dense star forming regions. More recent observations in regions with lower optical depth and observations at sub-mm wavelengths strongly suggest that 10\% would be a more appropriate choice for the maximum fractional polarization warm dust would have at the wavelengths of CMB experiments \citep{hil00}.

Direct optical polarimetry in regions of the sky a few degrees across at high latitudes confirms this prediction. For example, the region from $l = 275\degr - 295\degr, b = 60\degr - 70\degr$ shown in Figure 3 has optical polarimetry for over 25 stars in the compilation by \cite{hei96}. In this region we find a very uniform position angle $(\theta_{rms} = 0.25 rad)$. The observed fluctuations in fractional polarization $(P_{rms}/<P> = 0.4)$ are about a factor of two higher than expected given the observed $\theta_{rms}$. This is presumably due to fluctuations in optical depth across the region, not fluctuations in magnetic field geometry.

\begin{figure}[!ht]
\begin{center}
\includegraphics[scale=0.6]{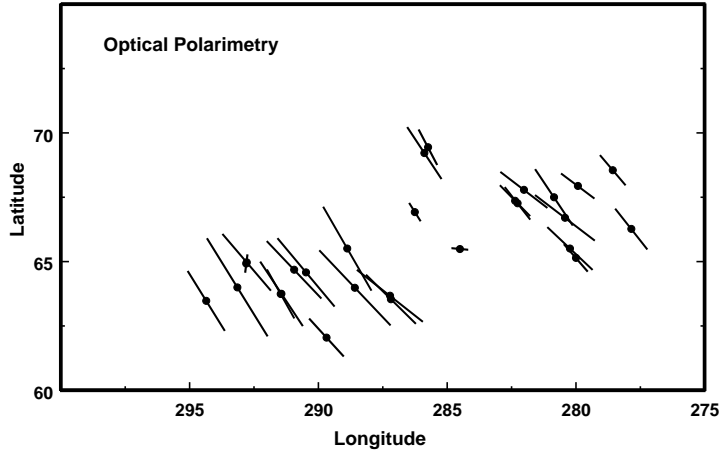}
\caption{\footnotesize Optical polarimetry of stars at high galactic latitude in a region taken from the compilation by \cite{hei96}. The measured dispersions in position angle and fractional polarization are $\theta_{rms} = 0.25 rad$ and $P_{rms}/<P> = 0.4$. Note how uniform the magnetic field geometry is over a span of several degrees.}
\end{center}
\end{figure}

\section{FIR Surface Brightness}

This leaves fluctuations in total intensity as the major source of fluctuations in Q and U from emission of aligned dust at high latitudes. Using the 100~\micron\ map from \cite{sch98} and the predicted average fractional polarization from Figure 2 for one decorrelation cell, we can predict the fluctuations in Q and U due to fluctuations in total intensity at FIR wavelengths. To make a prediction at a frequency of 150 GHz, we use the following input parameters: 

\begin{enumerate}

  \item Assume $<P_{e}>$ and $\theta$ are both constant with $<P_{e}> = 5\%$. 

  \item The measured fluctuations of the intensity in the central $40\degr \times 40\degr$ region of the 100~\micron\ map is $I_{rms} = 0.4$MJy/sr using 10' bins.
  
  \item Extrapolate the intensity from 100~\micron\ to 150 GHz using a power law index of 1.75 for the emissivity. 
  
  \item Arbitrarily assign Q to lie along the magnetic field direction. In this configuration the average value of U will be 0.
    
\end{enumerate}

\noindent With these parameters, we find that $Q_{rms} \sim 10^{-7}I_{CMB}$ due to fluctuations in total intensity alone. 

We can improve this estimate somewhat if we include the lesser effects of fluctuations in the magnetic field geometry. Taking the observed dispersion in position angle for the region of sky shown in Figure 3 $(\theta_{rms} = 0.25 rad)$ and using the JKD model to predict the fluctuations in $P_a$, we find $P_{rms}/<P> = 0.2$, about half the observed value. Combining the effects of fluctuations in intensity at 150 GHz with fluctuations in magnetic field geometry for this one piece of the sky, we obtain the predicted values for $Q_{rms}$ and $U_{rms}$ given in Table 1. 

\begin{table}[!ht]
\begin{center}
\begin{tabular}{ccc}
\multicolumn{3}{c}{Table 1. Predicted Contamination} \cr & & \\ \hline
GHz & $Q_{rms}/I_{CMB}$ & $U_{rms}/I_{CMB}$ \cr
 & ($10^{-7}$) & ($10^{-7}$)\\ \hline
 100 & 0.25 & 0.12 \\
 150 & 0.91 & 0.44 \\
 200 & 2.6 & 1.3 \\ \hline
\end{tabular}
\end{center}
\end{table}

\end{document}